# Optical Thermometry with Quantum Emitters in Hexagonal Boron Nitride

Yongliang Chen,[1] Thinh Ngoc Tran,[1] Ngoc My Hanh Duong,[1] Chi Li,[1] Milos Toth,[1,2] Carlo Bradac,[1] Igor Aharonovich,[1,2] Alexander Solntsev,[1] Toan Trong Tran[1*]

1. School of Mathematical and Physical Sciences, University of Technology Sydney, Ultimo, NSW, 2007, Australia

2. ARC Center of Excellence for Transformative Meta-Optical Systems (TMOS), Faculty of Science, University of Technology Sydney, Australia

*Corresponding author: trongtoan.tran@uts.edu.au

## Abstract

Nanoscale optical thermometry is a promising non-contact route for measuring local temperature with both high sensitivity and spatial resolution. In this work, we present a deterministic optical thermometry technique based on quantum emitters in nanoscale hexagonal boron-nitride. We show that these nanothermometers exhibit better performance than that of homologous, all-optical nanothermometers both in sensitivity and range of working temperature. We demonstrate their effectiveness as nanothermometers by monitoring the local temperature at specific locations in a variety of custom-built micro-circuits. This work opens new avenues for nanoscale temperature measurements and heat flow studies in miniaturized, integrated devices.

## Introduction

The ability to measure and control temperature at the nanoscale is critical to several areas of research,[1-3] from biology[4-5] and medicine[6-9] to solid-state-based, nano-electronics and -photonics[10-12]. At the nanoscale—where the Fourier's law no longer holds[13] heat transport behaves differently than in bulk materials. For instance, local fluctuations in sub-micron integrated circuits can lead to the formation of hot-spots, which often result in performance deterioration or even catastrophic failure. As a new generation of optoelectronic devices made from (stacks of heterogeneous) two-dimensional (2D) materials is on the rise,[14-17] the need to measure local temperatures with resolution down to few atomic layers is becoming imperative. The motivation is two-fold. The first is practical: to drive the design of devices performing at the cusp between top performance and critical failure. The second is fundamental: to explore the complex thermal dynamics at play when circuitry is miniaturized and become quasi-two-dimensional.[1-3]

Nanothermometry, based on non-invasive all-optical measurements, is well established.[1] Nanoscale thermal probes include fluorescent dyes,[18] organic polymers,[19] quantum dots,[20] gold nanoparticles and nanorods,[21] upconversion nanoparticles[22-23] and color centers in nanodiamonds.[3] However, only a small minority of these nanosensors can withstand the extreme temperatures at which high-power electronics operate, especially those reached in the hot-spots of sub-micron integrated circuits. Diamond-based nanothermometry is indeed a viable option[24-25] and includes, desirably, an extensive variety of methods based on mapping temperature changes onto the photoluminescence of different color centers—the nitrogen-vacancy (NV),[26-27] the silicon-vacancy (SiV),[28-29] the germanium-vacancy (GeV)[30-31] and the tin-vacancy (SnV) center.[32] Nevertheless, one of the main issues of diamond nanothermometers lies in their thermal contacts. While this is not a problem[29] when the nanodiamonds are entirely surrounded by the medium, e.g., in water or in the cytoplasm of a cell, thermal contact can be rather poor when nanodiamonds are merely deposited on a substrate.[33] Furthermore, deterministic and reliable placement of single nanoparticles in the near proximity of the object to be sensed can be laborious, albeit accurate. [34-36] In this work, we propose a nanothermometry technique that addresses these shortcomings.

The technique relies on color centers in layered two-dimensional (2D) hexagonal boron nitride (hBN). These are atom-like defects that show ultra-bright,[37] high-purity,[38] and photostable emission to up to 800 K[39] and whose zero-phonon line (ZPL) energy and linewidth display temperature dependence.[40] These hBN nanothermometers possess a few distinguishing features. *i)* Unlike other emitter-based nanothermometers which rely on large ensemble of colour centres to improve signal detection (and thus resolution), hBN nanothermometers can achieve comparable performances with single emitters thanks to their superior brightness. It follows that hBN nanosensors can be as small as just a few nm,[41] which is crucial for increasing spatial resolution. *ii)* Their sensitivity is comparable—in many cases higher (see below)—than homologous all-optical techniques. *iii)* They can reliably operate over a large temperature range, 0 K < ΔT ≲ 800 K. *iv)* The two-dimensional nature of hBN guarantees excellent thermal contact, which can be a limiting factor for other nanoparticle-based nanosensors. *v)* Furthermore, hBN-based nanothermometers are compatible—in fact they are ideal—for measuring temperatures and temperature gradients in 2D materials. Hexagonal boron-nitride is commonly used as a 'dielectric spacer' in many photo-electronic devices designed from stacks of heterogeneous 2D materials, which makes its potential use as a thermo-sensing layer an attractive possibility.[42]

The key concept of optical thermometry with hBN quantum emitters is illustrated in **Figure 1a**. A green laser (532-nm) is used to excite individual quantum emitters while their fluorescence is monitored. The emitters' photoluminescence zero-phonon line (ZPL) wavelength and linewidth depends on temperature due to lattice expansion and contraction of hBN, as well as defect-lattice phonon-mediated interactions.[40] **Figure 1b** shows the spectra of a typical hBN single photon emitter acquired at 296 K and 500 K and highlights a distinct temperature dependence. Specifically for this emitter, from 296 to 500 K, the ZPL wavelength red-shifts by ~8.5 nm and the ZPL linewidth broadens by ~13 nm, showing that both quantities can be used for thermal sensing. Notably, owing to the superb temperature stability of the material, hBN-based nanothermometers can operate at temperatures as high as 800 K without photobleaching or degradation—unlike other thermal probes such as fluorescent dyes, organic

polymers or colloidal quantum dots that are prone to oxidation or photobleaching at high temperatures.[2]

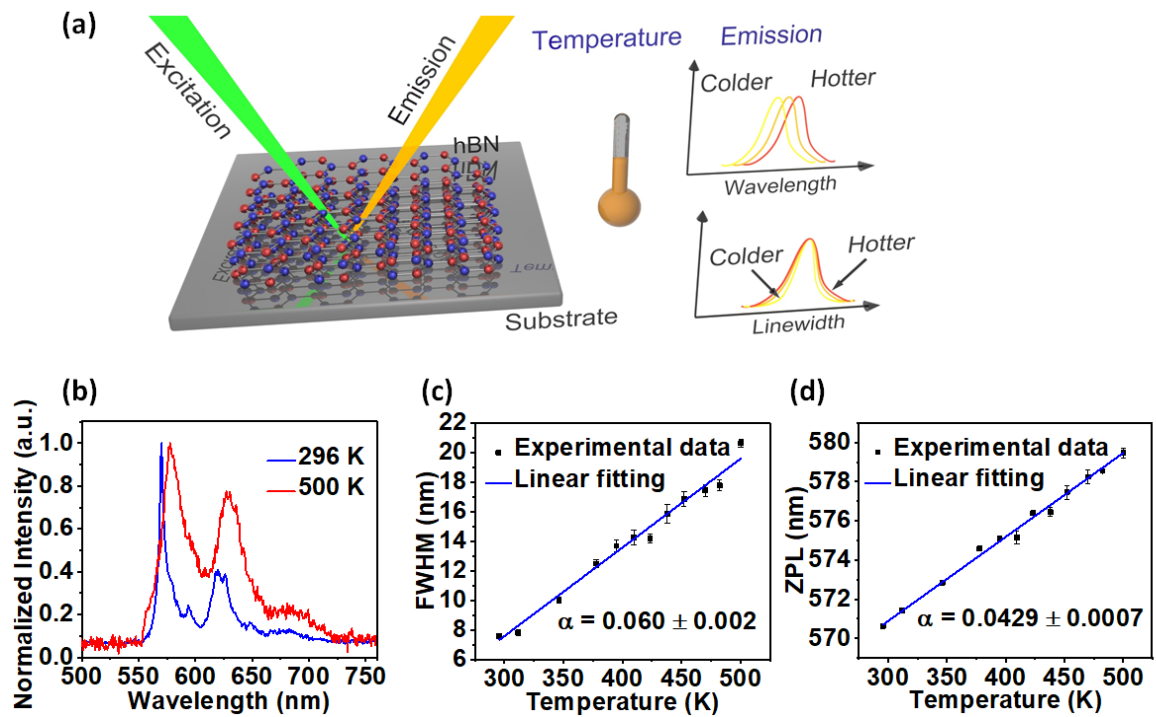

**Figure 1. a)** Schematic of optical thermometry using a quantum emitter in hexagonal boron nitride. Thermally-driven changes in the emitter's photoluminescence zero-phonon line (ZPL) wavelength and linewidth can be used to map the local temperatures of the substrate. **b)** Fluorescence spectra of a typical hBN nanothermometer where the ZPL wavelength and linewidth red-shifts and broadens, respectively, as temperature is increased from 296 K to 500 K. **c)** Variation of the ZPL's full width at half maximum (FWHM) as a function of temperature. The blue solid line is a linear fit to the experimental data (black squares). **d)** Variation of the ZPL barycentre wavelength as a function temperature. The blue solid line is a linear fit to the experimental data (black squares). Each data point in (c) and (d) was extracted by fitting a Lorentzian function to the ZPL. Error bars are calculated using the standard deviation for five consecutive 10-s spectral acquisitions performed at each value of temperature. A 20-min waiting time was allowed for thermal stabilization at each temperature.

**Figure 1** shows the basic performance of a typical hBN-based nanothermometer. The calibration was carried out using an external heater in contact with a thermocouple; the centre's ZPL and full width at half maximum (FWHM) were measured at twelve set temperatures over the range 296–500 K. Five photoluminescence (PL) spectra were captured and averaged at each temperature. A Lorentzian function was used to fit the ZPL position and linewidth: **Figure 1c** and **1d** show their respective temperature dependence. Both quantities scale linearly with temperature over the investigated range, which is consistent with previous report.[39] Linearity is a desirable—yet not required—feature in a nanothermometer as it grants constant sensitivity over the working temperature range.[1, 32] The extracted slopes from the linear fits are $(60 \pm 2) \times 10^{-3}$ nm·K$^{-1}$ and $(42.9 \pm 0.7) \times 10^{-3}$ nm·K$^{-1}$ for the ZPL linewidth and ZPL shifts, respectively. It should be noted that the observed linearity might not extend beyond the investigated range of temperatures. In fact, in semiconductors, temperature-driven variations of the bulk lattice constant commonly produce shifts for an intraband defect's ZPL that scale as $\sim T^3$. Simultaneously, the ZPL linewidth of hBN emitters is expected to broaden following a nearly exponential dependence with temperature due to the interaction of the defect's transition dipole with lattice acoustic phonons.[40] Note that this nearly-exponential dependence is different from that of other 3D semiconductors emitters whose temperature-driven linewidth broadening have been reported to follow $T^3$, $T^5$ and $T^7$ dependencies.[43-44] Based on these observations, for hBN emitters the precision in estimating the ZPL shift is higher than that for estimating the ZPL linewidth broadening, especially at high temperatures. This is due to the inability—exacerbated at high temperatures—to fit the ZPL linewidth with a single Lorentzian. Therefore, for the remainder of this work we utilize the emitter's shift in ZPL wavelength rather than its linewidth broadening to characterize the performance of hBN nanothermometers.

To demonstrate the utility of hBN-based nanothermometers we employed them to monitor the temperature of custom-designed micro-circuits. These circuits were patterned by electron beam lithography (EBL) with polymethyl methacrylate (PMMA) as the resist, followed by development with methyl isobutyl ketone (MIBK) to form the resist pattern. The metal deposition (gold, chromium or nickel) was performed before the lift-off process took place using warm acetone to remove metal and resist residues. A key feature of our thermo-sensing method is that we can monitor the temperature in specific points of interest of the circuit as the hBN nanoflakes can be positioned deterministically with respect to a target object. This is achieved via an align dry-transfer technique for two-dimensional (2D) materials that we modified (cf. Supporting Information **Figure S1**) from previous work[45-47] and allows for the positioning of the sensor with spatial accuracy <1 μm.

**Figure 2a** illustrates the transferring process of a single hBN nanoflake onto a targeted position of the microcircuit. We first fabricate a custom-made, dry stamp consisting of a polyvinyl alcohol (PVA)-coated polydimethylsiloxane (PDMS) hemisphere (3 mm in radius and 1 mm in thickness) by consecutive depositions of PDMS and PVA on a glass slide. The transparency of the PDMS/PVA stamp allows us to align its center to a selected hBN nanothermometer situated on a silicon substrate with the aid of a long-range objective lens mounted on a CMOS camera. The silicon substrate is then heated up to 70 °C and the stamp is brought into contact with the hBN nanothermometer. Since the temperature is close to the glass transition temperature of PVA (80 °C), the polymer is softened, which causes the hBN nanoflake to partially set in the PVA polymer. The hemispherical shape of the stamp allows for the selective pick-up of the hBN nanothermometer with an effective contact area of ~50 × 50 μm$^2$. Upon reducing the temperature to ambient conditions, the PVA-plus-hBN stamp solidifies. It is lifted and is ready to be moved over the target sample circuit, which is treated beforehand in a commercial UV ozone cleaner for 5 min to remove surface contaminants and render the surface highly hydrophilic. The stamp-plus-hBN flake is aligned to the target location of the circuit, brought into contact, and heated at 110 °C at which the PVA liquifies releasing the hBN nanothermometer; the temperature is kept at 110 °C for a minute to ensure complete release. The stamp is then lifted from the sample, leaving the hBN nanothermometer on the circuit. The PVA layer is easily washed off with warm water. Notably, our method is universal and can be applied for a wide range of target materials possessing different surface energies. To demonstrate this point, we will perform the deterministic transfers on circuits made from three representative metals, namely gold, chromium and nickel. The details will be discussed in the later section of the text.

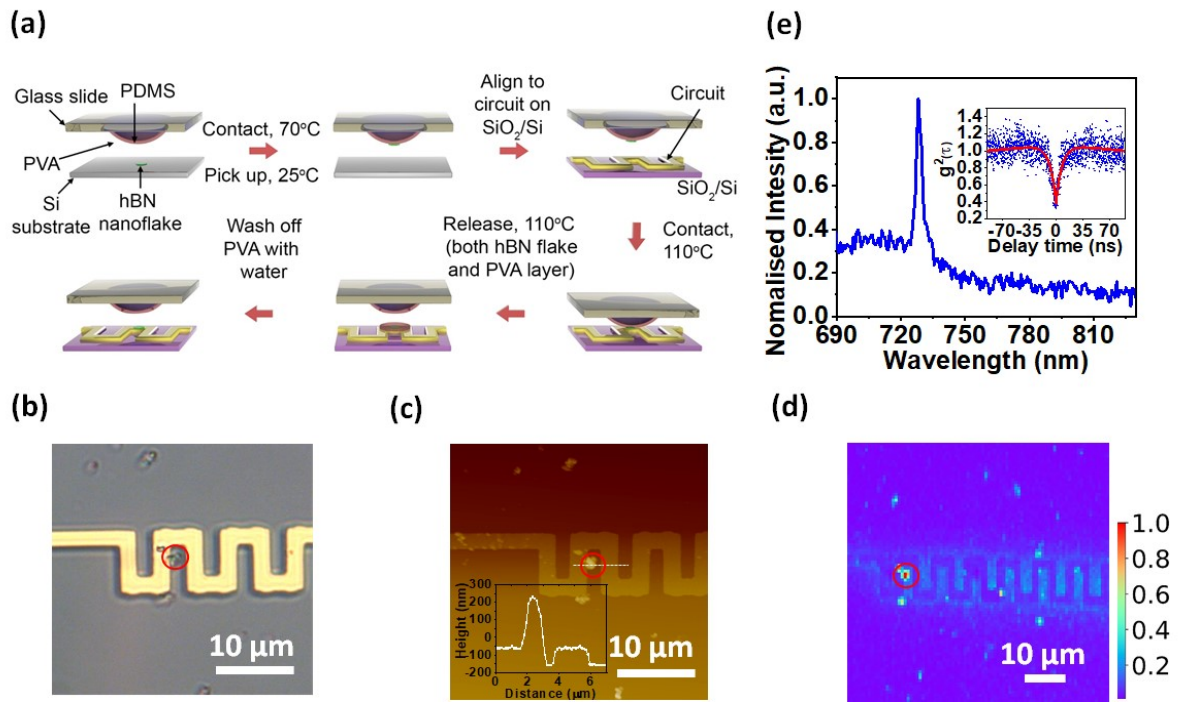

**Figure 2.** Deterministic transfer and characterization of the transferred hBN nanothermometer. **a)** Schematic of the dry-transfer process developed to transfer any hBN nanoflake from a $SiO_2$ substrate (where the hBN nanoflakes are pre-characterized) onto the targeted location of a microcircuit (see main text). **b)** Optical image of the hBN nanoflake (marked by the red circle) transferred onto the circuit. **c)** Corresponding AFM image and height profile (inset) of the hBN nanoflake onto the circuit; the profile data refers to the section indicated by the white dashed line. **d)** confocal PL map showing that the emitter stays optically active after transfer. **e)** PL spectrum from the quantum emitter after transfer. Inset: second-order autocorrelation function showing $g^{(2)}(\tau = 0) = 0.38$, which is indicative of a single-photon emitter.

**Figure 2b** shows a wide-field optical image with a typical hBN flake-nanothermometer (circled in red) successfully transferred onto the circuit. The flake was pre-characterised prior to transfer to confirm it hosted a stable single-photon emitter. After transfer, Raman spectroscopy confirmed that the nanothermometer was high-quality hBN, with the $E_{2g}$ peak appearing at ~1365 cm$^{-1}$ with a full width half maximum (FWHM) of ~10 cm$^{-1}$ (cf. Supporting Information **Figure S2a**). We used atomic force microscopy (AFM) to determine the height and dimensions of the nanothermometer (**Figure 2c**): the lateral and vertical dimensions of the hBN flake are ~900 nm and ~280 nm, respectively. We verified that the hBN emitter was still optically active after transfer via confocal microscopy (**Figure 2d**). **Figure 2e** shows the spectrum and second order autocorrelation function, $g^{(2)}(\tau)$, collected for the emitter after transfer onto the circuit. The value of $g^{(2)}(\tau) < 0.5$ indicates that the photoluminescence originates from a hBN single photon emitter (SPE). Time-resolved photoluminescence and polarization spectroscopy measurements also show that the emitter possesses a lifetime of (6.1 ± 0.1) ns and linearly polarised emission (cf. Supporting Information **Figure S2b** and **S2c**).

Note that using a single emitter rather than an ensemble offers a series of advantages. Besides the aforementioned and obvious benefit of the reduced size of the nanosensor, working with a single emitter should also reduce the error on the measured observables—ZPL position and linewidth in this case. The photoluminescence spectrum of a single emitter, for instance, is not subject to inhomogeneous broadening; and whilst an individual emitter can still be affected by local fluctuations, e.g. strain, nearby trapped charges and other lattice defects these can be more easily accounted for when selecting the hBN candidate nanothermometer prior to pick-up.

**Figures 3a** and **3b** show the schematic and the actual picture of the circuit we used to conduct our thermosensing experiments. We gradually increased the current flown through the circuit to create realistic working conditions. Upon flowing current through it, the circuit is heated up by the Joule effect. Each value of the current was held for 10 minutes to guarantee thermal equilibrium was reached, after which five consecutive 10-s spectra were acquired. Each spectrum was fitted with a Lorentzian, and the corresponding ZPL position was extracted. **Figure 3c** shows a plot of the shift in ZPL position as a function of the applied current. From 10 to 50 mA, the ZPL wavelength red-shifts by about ~0.25 nm, while from 50 to 90 mA it red-shifts by ~2.25 nm. By utilizing the temperature calibration curve for the ZPL position we then obtained an actual temperature reading from the circuit as a function of input current (**Figure 3d**). Specifically, we determined that the temperature increased from ~300 K to ~310 K as the current varied from 10 to 50 mA, and from ~310 K to ~360 K as the current was increased from 50 to 90 mA. Similar results (data not shown) were obtained when monitoring the change of the ZPL linewidth but, as discussed above, the linewidth-based measurement is more prone to error and leads to lower resolution (cf. supporting information **Figure S4**). **Figure S3** in the Supplementary Information also shows how the resolution of the nanothermometer can be improved at the cost of bandwidth, i.e. improving the signal to noise ratio of the measurement by integrating for a longer time.

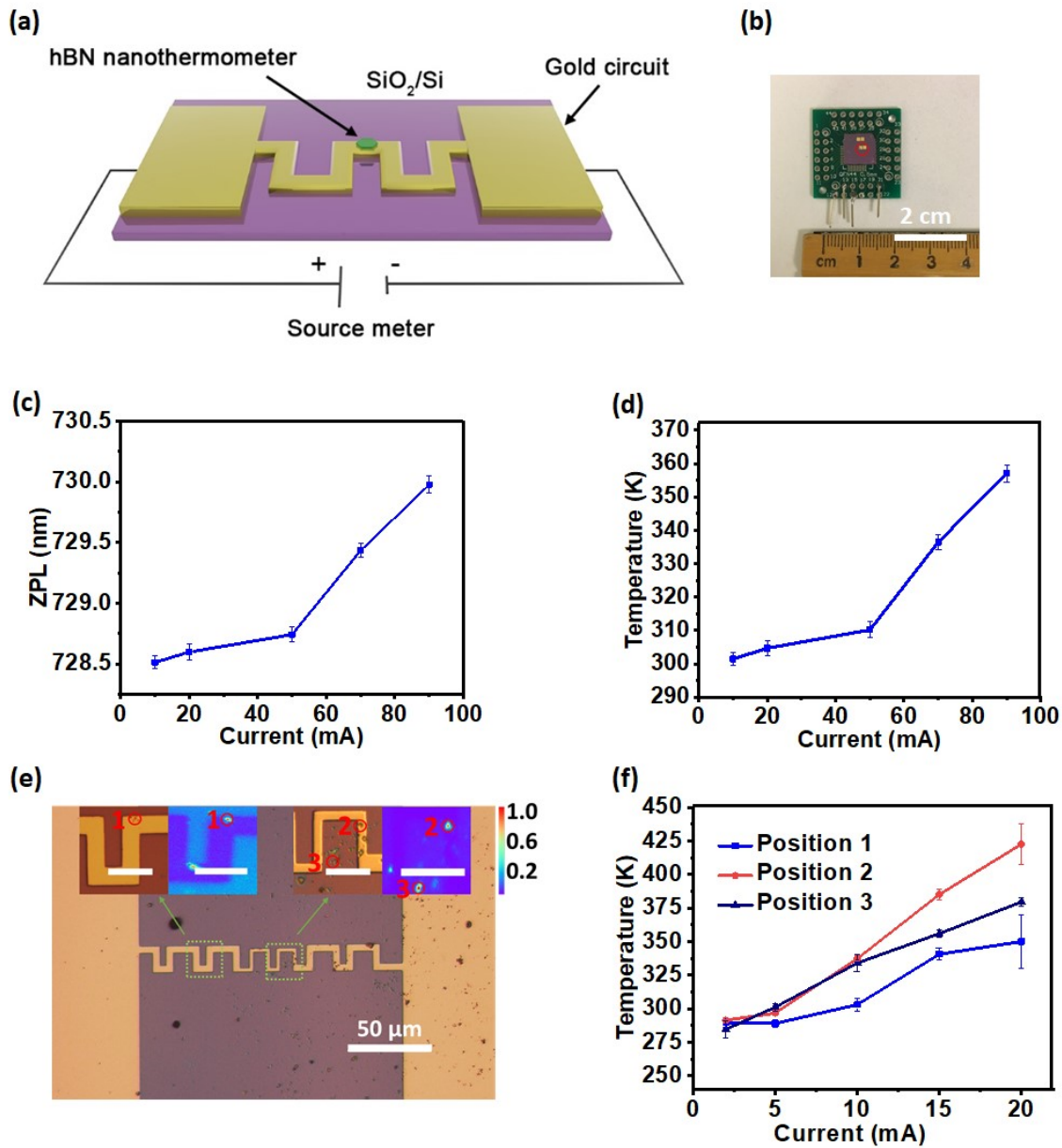

**Figure 3.** A real-time temperature measurement in micro-circuits using quantum emitters in transferred hBN nanoflakes. **a)** Schematic illustration of the microcircuit and hBN nanothermometer placed onto it for temperature mapping. **b)** Optical image of a gold microcircuit made by the EBL technique and used (red circle) in our first experiment. **c)** Shift of the quantum emitter's ZPL wavelength as an increasing current is applied. **d)** Corresponding values for the temperature as estimated by the ZPL shift after the emitter has been calibrated

against a known set of temperatures. **e)** Optical image of the chromium microcircuit made by EBL and used in our second experiment. Insets: zoomed-in optical images of portions of the microcircuit where hBN nanoflakes have been transferred on (numbered circles 1–3) and corresponding PL confocal map. The inset scale is 10 um. **f)** Temperature measurement for the three positions in (e) as current through the circuit is increased. Temperatures are estimated from the calibrated shift in the emitters' ZPL.

To further demonstrate the utility of the technique in practical scenarios, we monitored the temperature of multiple target locations on a chromium micro-circuit by measuring the photoluminescence of three hBN nanoflakes transferred onto it (**Figure 3e**). The metallic contacts were fabricated with different widths in the inner (2-μm width) and outer (4-μm width) part of the circuit. Given the different resistivity and heat dissipation rates of each of the three target positions, their respective temperature was expected to be different.

Each hBN-nanothermometer was a single quantum emitter. Their photoluminescence and temperature calibration is shown in the Supplementary Information, Figures S4g (Position 1), S4e (Position 2), and S4d (Position 3). **Figure 3f** shows the estimated temperature of each position as the current was increased from 2 mA to 20 mA. As expected, the temperature in position 3 was consistently lower than that in positions 1 and 2 as in 3 the hBN-nanothermometer was not in contact with the metal (it was ~3 μm away). The temperature was the highest in position 2 where the circuit is the narrowest. **Figure S5** in the Supplementary Information shows the details of the measurement. Briefly, both position 1 and 2 (on-circuit) exhibited a linear temperature vs power relationship whereas in position 3 (off-circuit) the relationship was non-linear. While the linear relationship with the temperature of on-circuits locations is due to the simple Joule heating phenomenon, the non-linear trend at the off-circuit location is attributed to the heat sink effect by the substrate, consistent with previous studies.[48-49] The Supplementary Information (**Figure S6**) shows a third example with a hBN nanothermometer transferred onto a nickel circuit and used to map its temperature (**Fig. S4h** and **S6c**).

We finally turn our attention to comparing the performance of our hBN nanothermometers to that of other homologous nanoscale thermal probes. The comparison is conducted only considering other nanothermometers which are, like ours, all-optical and specifically map temperature values onto spectral shifts in photoluminescence. **Figure 4** summarizes the analysis by showing the sensitivity of each technique (y-axis) as well as the maximum working temperature (x-axis). Nanothermometers in the top-right corner of the table have, desirably, both high sensitivity and large working temperature range. The sensitivity of the hBN nanothermometers presented in this work (cf. Supplementary Information, **Fig. S7**) is, comparatively, almost three-fold and five-fold higher than that of nanothermometers based on diamond tin-vacancy[32] and germanium-vacancy colour centres[50] Quantum-dot nanothermometers can reach higher sensitivities, yet their narrow temperature range can prevent their application in realizations requiring the monitoring of high temperatures.[51]

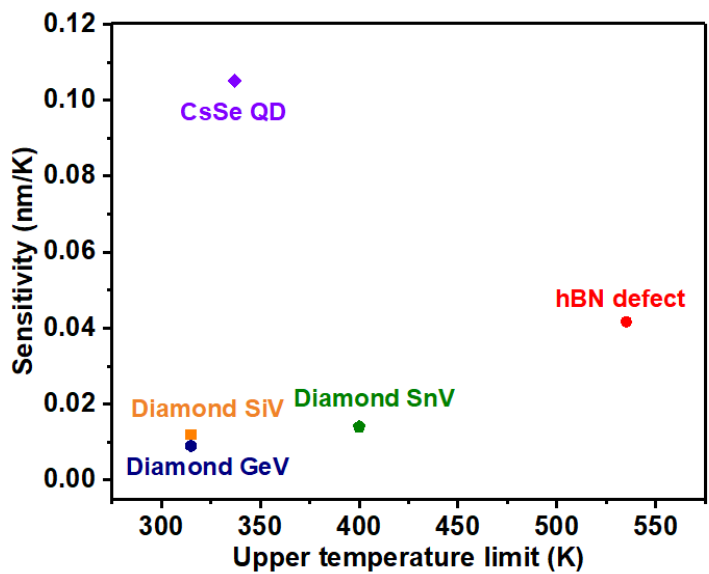

**Figure 4.** Sensitivity (y-axis) and upper-temperature limit (x-axis), while maintaining linear response, of some of the common nanothermometers which use shifts in ZPL wavelength centre to map temperature.

In conclusion, we have demonstrated a novel nanothermometry technique based on measuring photoluminescence spectral shifts of bright single photon sources in hexagonal boron-nitride. The technique has several advantages over homologous all-optical thermosensing methods. The single nature of the emitter results in smaller nanosensor sizes (thus potentially better spatial resolution), while still achieving better sensitivities over a large temperature range, 0 K $< \Delta T \lesssim 800$ K. Furthermore, the 2D nature of hBN grants these nanosensors excellent thermal contact and compatibility with the measurement of temperatures in stack of heterogeneous 2D materials. We have also shown the versatility of the technique by measuring the temperature of selected target points in a series of custom-built micro-circuits made of gold, chromium and nickel. Our results offer an intriguing alternative to other nanothermometry methods especially in the context of characterizing the performance of high-power electronics as well as investigating the complex thermal dynamics of miniaturized, quasi-two-dimensional electronic devices.

## Experimental Methods

**Preparation of hBN nanoflakes on silicon (Si) substrate.** The commercial solution of 200-nm hBN nanoflakes (Graphene Supermarket) is dropcast onto a Si substrate. The substrate is then annealed for 1 hour in vacuum (5 mTorr) in a tube furnace (Lindberg/ Blue Minimite) at 1000 °C with Argon flow rate of 50 sccm, which is used to activate and stabilize the hBN quantum emitters.

**Preparation of the microcircuit on the $SiO_2$/Si substrate by electron beam lithography (EBL).** An amount of ~0.1 ml of 950 PMMA is dropcast on the thermally grown $SiO_2$ (~300 nm) on Si substrate and spun for 1 min at 3000 rpm to get an homogeneous PMMA coating (thickness ~200 nm). Microcircuits are patterned using a conventional SEM (Zeiss Supra 55VP) equipped with the Raith EBL system. The pattern PMMA is removed via immersion into MIBK/IPS (1:3) solution. The substrate is then coated with a metal (gold, chromium or nickel) film (thickness ~60 nm) using a lab-built plasma sputter deposition chamber. After immersion in high-purity warm acetone, the metal parts on the PMMA are washed off together with the PMMA itself leaving behind the patterned metal microcircuits (thickness ~60 nm).

**Align transfer.** The deterministic transfer was conducted by a lab-built align-transfer setup which includes a conventional optical microscope equipped with a long working distance objective, a digital camera connected to a computer and two micropositioners (one controlling the sample and the other the stamp) as shown in **Figure S1** (cf. also main text). The optical microscope is used to do the alignment. The stamp is used to pick up and release a selected hBN nanoflake. The stamp is made by a mixture, 10:1 volume ratio, of Sylgard 184 prepolymer and crosslinking agent to form a hemispherical PDMS base on the slide; this is followed by dropping a PVA solution to form a PVA coating. The heating stage is connected to a source meter unit (SMU, Keithley 2400), while a thermocouple is used to track the temperature of the sample.

**Optical characterization.** The substrates and microcircuits with hBN nanoflakes are mounted onto a three-dimensional micropositioner in a lab-built photoluminescence (PL) confocal microscopy setup. The system is equipped with a 532-nm continuous wave (Gem 532, Laser quantum Ltd.) laser and a long working distance air objective (numerical aperture 0.7, Olympus). The emitted photons are collected through a graded-index multimode fiber with an aperture of 62.5 μm. A 20% / 80% fiber-splitter directs the photons (20%) to an avalanche photodiode (APD) (Excelitas Technologies) and (80%) to a spectrometer (QEPro, Ocean Optics). In the experiments hBN emitters are pre-characterised, with the samples being mounted on the heating stage connected to the SMU (for heating) and the thermocouple (for measuring the sample temperature). Second-order autocorrelation measurements are conducted in a Hanbury Brown-Twiss interferometer equipped with two APDs and a time-correlated single-photon counting module (PicoHarp 300, PicoQuant). Raman spectroscopy is carried out with a Renishaw in Via Raman$^{TM}$ microscope equipped with a 633-nm laser source. Time-resolved fluorescence measurements are performed using picosecond pulses at 510 nm, pulse width 100 ps and repetition rate 80 MHz (PiL051XTM, Advanced Laser Diode Systems A.L.S. GmbH).

## Author Information

### Corresponding author

*Email (T. T. Tran): trongtoan.tran@uts.edu.au

## Notes

The authors declare no competing financial interest

## Acknowledgments

We thank Dr. Sejeong Kim for building the transfer setup. We acknowledge financial support from the Australian Research Council (via DP180100077, DE180100810 and DP190101058).